\documentclass{ifacconf}  

\usepackage{epsfig} 
\usepackage{amsmath,amssymb,amsfonts} 

\usepackage{subfig}
\usepackage{float, color}
\usepackage{enumitem}

\usepackage{subfloat}
\usepackage{theorem}
\theoremstyle{plain} \theorembodyfont{\upshape}
\newtheorem{problem}{Problem}

\usepackage{natbib}  

\allowdisplaybreaks[1]

\newenvironment{varalgorithm}[1]
  {\algorithm\renewcommand{\thealgorithm}{#1}}
  {\endalgorithm}

\newenvironment{list4}{
	\begin{list}{$\bullet$}{%
			\setlength{\itemsep}{0.05cm}
			\setlength{\labelsep}{0.2cm}
			\setlength{\labelwidth}{0.3cm}
			\setlength{\parsep}{0in} 
			\setlength{\parskip}{0in}
			\setlength{\topsep}{0in} 
			\setlength{\partopsep}{0in}
			\setlength{\leftmargin}{0.16in}}}
	{\end{list}}

\newenvironment{list4a}{
	\begin{list}{$\bullet$}{%
			\setlength{\itemsep}{0.05cm}
			\setlength{\labelsep}{0.2cm}
			\setlength{\labelwidth}{0.3cm}
			\setlength{\parsep}{0in} 
			\setlength{\parskip}{0in}
			\setlength{\topsep}{0in} 
			\setlength{\partopsep}{0in}
			\setlength{\leftmargin}{0.16in}}}
	{\end{list}}

\usepackage{mathtools}

\begin{document}

 \DeclarePairedDelimiter{\floor}{\lfloor}{\rfloor}
  \DeclarePairedDelimiter{\ceil}{\lceil}{\rceil}
\thispagestyle{empty}
\pagestyle{plain}

\begin{frontmatter}

\title{
Distributed Optimal Allocation with Quantized Communication and Privacy-Preserving Guarantees}

\thanks[footnoteinfo]{%
This work was supported in part by   a  Distinguished Professor Grant from the Swedish Research Council (Org: JRL, project no: 3058).}

\author[First]{Jakob Nylöf}
\author[First]{Apostolos I. Rikos}
\author[Second]{Sebin Gracy}
\author[First]{and Karl Henrik Johansson}

\address[First]{
 Division of Decision and Control Systems, School of Electrical Engineering and Computer Science, KTH Royal Institute of Technology, and Digital Futures, Stockholm, Sweden. jnylof@kth.se, rikos@kth.se, kallej@kth.se}
 \address[Second]{Department of Electrical and Computer Engineering, Rice University, Houston, TX, USA. sebin.gracy@rice.edu}


\begin{abstract}
In this paper, we analyze the problem of optimally allocating resources in a distributed and privacy-preserving manner. 
We propose a novel distributed optimal resource allocation algorithm with privacy-preserving guarantees, which operates over a directed communication network. 
Our algorithm converges in finite time and allows each node to process and transmit quantized messages. 
Our algorithm utilizes a distributed quantized average consensus strategy combined with a privacy-preserving mechanism. 
We show that the algorithm converges in finite-time, and we prove that, under specific conditions on the network topology, nodes are able to preserve the privacy of their initial state. 
Finally, to illustrate the results, we consider an example where test kits need to be optimally allocated proportionally to the number of infections in a region. 
It is shown that the proposed privacy-preserving resource allocation algorithm performs well with an appropriate convergence rate under privacy guarantees.
\end{abstract}
\end{frontmatter}

\section{Introduction}
\label{sec:intro} 
In distributed systems and networks, various components (nodes) are often required to allocate a set of resources in an optimal way such that specific performance objectives are satisfied. 
Distributed optimal resource allocation is an optimization problem, and has many applications such as optimally scheduling tasks for data centers \citep{rikos2021optimal}, optimally coordinating the response of a set of distributed energy resources \citep{2020:Madi_Hadj_Garcia}, optimally allocating vaccines/tests for pandemic stabilization \citep{alex:vaccine}. 
Note that in the current literature there exist a variety of centralized algorithms for addressing optimal resource allocation problems (see for instance \citep{2013:Fang, 2012:Moghaddas}). 
However, a central entity could possibly suffer from processing issues due to network scale, and may also impose privacy risks (due to the gathering of all available data to a central entity). 
For these reasons, we aim to address the optimal resource allocation problem in a distributed fashion. 

Distributed optimization algorithms have received great attention recently, due to the wide variety of applications which range from distributed estimation to machine learning \citep{2018:Nedic_Olshevsky_Rabat, Tao:2019Survey}. 
However, a vast majority of algorithms in the current literature assume that the messages exchanged among nodes consist of real values with infinite precision (see, for instance, \citep{2020:Themis_Kalyvianaki, PreciadoTCNS14, ramirez2018distributed, 2014:Nedic_Beck}) and they exhibit asymptotic convergence within some error (see \citep{DOMINGUEZGARCIA201533}). 
Furthermore, most algorithms typically do not provide privacy-persevering guarantees (see \citep{2020:Madi_Hadj_Garcia, rikos2021privacy}). 
In this paper, we aim to address both of these issues, since in the current literature there is a need for finite-time distributed optimal resource allocation algorithms with \emph{privacy-preserving} guarantees and efficient communication, which exhibit finite time convergence. 
To illustrate the efficacy of our proposed algorithm, we consider the setting where vaccines (i.e., devices for testing whether a person is infected from a specific virus) have to be distributed in an optimal fashion 
over a network of cities dealing with an epidemic outbreak. 



\vspace{.2cm}

\noindent
\textbf{Main Contributions.} 
Our main contributions are as follows.
\begin{itemize}
    \item We present an optimal allocation algorithm with quantized communication and privacy-preservation guarantees; see Algorithm \ref{algorithm1}. 
    Furthermore, during the operation of our algorithm, each node terminates its operation once convergence has been achieved. 
    Note that it is the first distributed stopping mechanism adjusted to the algorithm's necessary privacy-preservation guarantees. 
    Our algorithm's operation is applied to distributed optimal test kit allocation problem over strongly connected networks.
    \item We analyze the convergence of Algorithm \ref{algorithm1}, and we show that all nodes calculate the optimal allocation in finite time with high probability; see Theorem~\ref{thm:test_allocation_private}. 
    \item We provide sufficient topological conditions for privacy-preservation of Algorithm~\ref{algorithm1}; see Theorem~\ref{thm:privacy_preservation}.
\end{itemize}

The optimal allocation algorithm in this paper uses properties of quantized average consensus algorithms \citep{rikos2021fast,aysal2008QAC,amini2019QAC,zhang2020QAC,lavaei2012QAC,kashyap2007QAC,chamie2016QAC} that allow nodes to exchange quantized messages. 
Transmissions of quantized messages are preformed asynchronously under a set of event-triggered conditions, which increase the efficiency of communication. 
Additionally, our algorithm is also able to guarantee privacy preservation of each node's initial state. 
The case of privacy preservation has been studied previously in \citep{hadjicostis2020privacy,rikos2021privacy,wang2019privacy,4448699,manitara2013QAC,GUPTA20179515}. 
In particular, \citep{rikos2021privacy} utilizes the injection of random quantized offsets into interaction messages transmitted from private nodes. 
However, the injection of quantized offsets is done in a deterministic manner. 
In contrast, in our paper the privacy preserving strategy is adjusted to the randomized nature of the quantized average consensus algorithm, as the injection of quantized offsets is performed according to a set of event-triggered conditions.

\section{NOTATION AND BACKGROUND}\label{sec:notation}

The sets of real numbers, positive real numbers, integers and natural numbers 
are denoted by $\mathbb{R},\; \mathbb{R}_+,\; \mathbb{Z}$ and $\mathbb{N}$, respectively.
For any $a\in \mathbb{R}$, the floor is defined as $\lfloor a \rfloor = \{\text{sup } b \in \mathbb{Z} \ | \ b\leq a \}$ 
and the ceiling as $\lceil a \rceil = \{\text{inf } b \in \mathbb{Z} \ | \ b\geq a \}$. 

\vspace{.2cm}

\noindent
\textbf{Graph-Theoretic Notions.}
The communication network is represented by a strongly connected directed graph (digraph) $\mathcal{G}_d = \mathcal{(V, E)}$ of $n$ nodes. 
In digraph $\mathcal{G}_d$, $\mathcal{V} =  \{v_1, v_2, \dots, v_n\}$ is the set of nodes, whose cardinality is denoted as $n  = | \mathcal{V} | \geq 2 $, and $\mathcal{E} \subseteq \mathcal{V} \times \mathcal{V} - \{ (v_j, v_j) \ | \ v_j \in \mathcal{V} \}$ is the set of edges (self-edges excluded) whose cardinality is denoted as $m = | \mathcal{E} |$. 
We assume that the given digraph $\mathcal{G}_d = (\mathcal{V}, \mathcal{E})$ is \textit{strongly connected} (i.e., for each pair of nodes $v_j, v_i \in \mathcal{V}$, $v_j \neq v_i$, there exists a directed path from $v_i$ to $v_j$). 
The diameter $D$ of a digraph is the longest shortest path between any two nodes $v_j, v_i \in \mathcal{V}$ in the network. 
The set of in-neighbors of $v_j$ is represented by $\mathcal{N}_j^- = \{ v_i \in \mathcal{V} \; | \; (v_j,v_i)\in \mathcal{E}\}$, and it is the subset of nodes that can directly transmit information to node $v_j$ is called. 
The \textit{in-degree} of $v_j$ and is denoted by $\mathcal{D}_j^- = | \mathcal{N}_j^- |$. 
The set of out-neighbors of $v_j$ is represented by $\mathcal{N}_j^+ = \{ v_l \in \mathcal{V} \; | \; (v_l,v_j)\in \mathcal{E}\}$, and it is the subset of nodes that can directly receive information from node $v_j$. 
The \textit{out-degree} of $v_j$ is denoted by $\mathcal{D}_j^+ = | \mathcal{N}_j^+ |$. 


\section{Problem Formulation}\label{probform_prelim}

\subsection{Distributed Optimal Resource Allocation Problem}\label{subsec_distr_alloc}

We state the following optimization problem, which is inspired by \citep{rikos2021optimal}. 
For each node $v_j \in \mathcal{V}$, we define the scalar quadratic local cost function $f_j : \mathbb{R} \mapsto \mathbb{R}$ as
\begin{equation}\label{local_cost_functions}
    f_j(z) = \dfrac{1}{2} \alpha_j (z - \chi_j)^2 , 
\end{equation}
where $\alpha_j \in \mathbb{R}_+$, 
$\chi_j \in \mathbb{R}_+$ is the demand at node $v_j$, and $z$ the global optimization parameter. 
In \eqref{local_cost_functions} we capture the cost of the node $v_j$ agreeing to obtain the quantity $z$ in relation to its demand $\chi_j$, where the weight $\alpha_j$ scales the cost.

The global cost function is the sum of the local cost functions \eqref{local_cost_functions} corresponding to each node $v_j \in \mathcal{V}$. 
The global cost function is the total cost of all nodes in the network agreeing to obtain the parameter $z$. 
Consequently, each node $v_j$ aims to obtain a value $z^*$ which minimizes the global cost function 
\begin{align}\label{opt:1}
z^* =  \arg\min_{z\in \mathcal{Z}} \sum\textstyle_{v_{i} \in \mathcal{V}} f_i(z) , 
\end{align}
where $\mathcal{Z}$ is the set of feasible values of parameter $z$. 
Equation \eqref{opt:1} has a closed form solution given by
\begin{align}\label{eq:closedform}
z^* =  \frac{\sum_{v_{i} \in \mathcal{V}} \alpha_i \chi_{i}}{\sum_{v_{i} \in \mathcal{V}} \alpha_i}.
\end{align}
Note that if $\alpha_i =1$ for all $v_{i}\in\mathcal{V}$, then the solution is the average of the initial states. 

\subsection{Modification of Optimal Resource Allocation Problem}\label{subsec_modific_alloc}


Consider an optimization step $m$ which represents a day on which we aim to find an optimal allocation of test kits to number of infections. 
For every node $v_j \in \mathcal{V}$, denote the local number of stored test kits by $u_j[m]$, received test kits by $l_j[m]$, and number of infections by $\lambda_j[m]$. 
Note here that these quantities are positive integers (which enables efficient communication since they are quantized values). 
Define $w_j^*[m]$ as the number of test kits added (or, if negative, subtracted) to the stored test kits in order to achieve the optimal allocation of the available test kits. 
We refer to $w_j^*[m]$ as the optimal allocation. 
Furthermore, denote the global number of stored test kits by $u_{tot}[m] = \sum_{v_i \in \mathcal{V}}u_i[m]$, global number of received test kits by $l_{tot}[m] = \sum_{v_i \in \mathcal{V}}l_i[m]$ and global number of infections by $\lambda_{tot}[m] = \sum_{v_i \in \mathcal{V}}\lambda_i[m]$. 
We drop the index $m$ in the sequel (since we aim to find the optimal allocation of test kits in the same way during each optimization step).
We now state following problem \textbf{P1.} which will be used as a framework in order to formulate the problem of interest in this paper (defined as Problem~\ref{prob:1} at the end of this section). 

\noindent
\textbf{P1.}
Formulate a distributed algorithm that allows each node $v_j$ to calculate the optimal allocation $w_j^*$ so that its local ratio of test kits to number of infections equals the global ratio of test kits to number of infections in the entire network. 

To solve \textbf{P1.}, we aim to find $w_j^*$ such that 
\begin{align}\label{cond:balance}
&\frac{w_j^* + u_j}{\lambda_j} = q, \ \forall v_{j} \in \mathcal{V}\\
&\text{where} \quad q = \frac{l_{\mathrm{tot}} + u_{\mathrm{tot}}}{\lambda_{\mathrm{tot}}}.\label{eq:global_average}
\end{align}

\noindent Note that $q = (\sum_{v_{i} \in \mathcal{V}} \lambda_i \frac{l_{i}+u_{i}}{\lambda_i})/(\sum_{v_{i} \in \mathcal{V}} \lambda_i)$ is the same as \eqref{eq:closedform} with $\alpha_j = \lambda_j$, and $\chi_j = (l_j + u_j)/\lambda_j$ for all $v_j \in \mathcal{V}$. 
Equation~\eqref{cond:balance} thus implies that $(w_j^* + u_j)/\lambda_j$ is the solution to the optimization problem \eqref{opt:1} where the weight $\alpha_j$ is the number of infections and $\chi_j$ the initial test kits to number of infections located at every node. 
Hence, we require every node to calculate the global test kits to number of infections given by \eqref{eq:global_average} and then solve for $w_j^*$ in \eqref{cond:balance}. 
The quantized coordination algorithm considered in this paper \citep{rikos2021fast} allows each node to calculate either the ceiling or the floor of $q$ which yields the optimal allocation 
\begin{equation}
    \label{eq:quantized_optimal_workload}
    w_j^* = \ceil*{q}\lambda_j -u_j \text{ or } \floor*{q}\lambda_j -u_j,\ \forall v_j \in \mathcal{V}.
\end{equation}
Equation \eqref{eq:quantized_optimal_workload} may introduce a larger quantization error compared to solving for $w_j^*$ in \eqref{cond:balance}. 
However, the event-triggering operation and the exchange of integer-valued messages increases the efficiency of communication while it maintains a fast convergence speed.

\subsection{Distributed Privacy-Preserving Optimal Resource \\ Allocation Problem}\label{Opt_Prob_Form} 

\begin{figure}[h] 
\begin{center} 
\includegraphics[width=0.5\columnwidth]{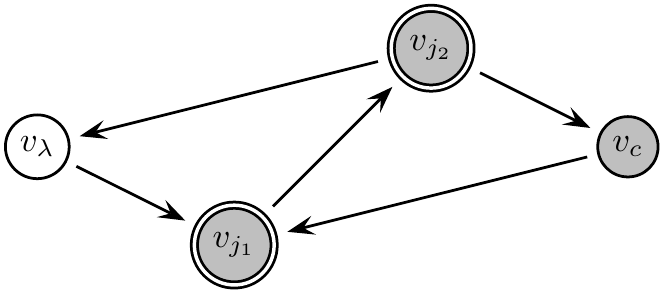}
\caption{Example of a digraph with the different types of nodes in the network: nodes $v_{j_1}, v_{j_2} \in \mathcal{V}_p$ that wish to preserve their privacy, node $v_{c} \in \mathcal{V}_c$ that is curious (wishes to identify the initial states of other nodes), and node $v_{\lambda} \in \mathcal{V}_{n}$ that is neither curious nor wishes to preserve its privacy\vspace{-0.32cm}.}
\label{prob_form_graph}
\end{center}
\end{figure}

The problem we present in this paper is denoted as Problem~\ref{prob:1}.
It is borrowed from \citep{rikos2021privacy} and it is adjusted to the optimal allocation scenario. 
Consider a strongly connected digraph $\mathcal{G}_d = (\mathcal{V}, \mathcal{E})$, where $| \mathcal{V} | \geq 3$. 
The node set $\mathcal{V}$ is partitioned into three subsets: i) a subset of nodes $v_j \in \mathcal{V}_p \subset \mathcal{V}$ that wish to preserve their privacy by not revealing their initial states to other nodes, ii) the subset of nodes $v_c \in \mathcal{V}_c \subset \mathcal{V}$ that are curious (i.e., thety try to identify the initial states of all or a subset of nodes in the network and they are possibly colluding among themselves), and iii) the subset of nodes $v_i \in \mathcal{V}_n \subset \mathcal{V}$ that are neutral (i.e., they neither wish to preserve their privacy nor identify the initial states of other nodes). 
An example is shown in Fig.~\ref{prob_form_graph} (borrowed from  \citep{rikos2021privacy}). 

We now provide below an analytical definition of the concept of privacy. 

\begin{defn}\label{Definition_Quant_Privacy}
A node $v_j \in \mathcal{V}_p$ is said to preserve the privacy of its initial state (denoted as $y_j[0] \in \mathbb{Z}$) if the value $y_j[0]$ cannot be inferred by the curious nodes in $\mathcal{V}_c$ at any point during the operation of the protocol. 
More specifically, the curious nodes can only determine a range $[\alpha,\beta]$ ($\alpha < \beta$) in which the values $y_j[0]$ lie in, and $v_j \in \mathcal{V}_p$ can make $\alpha \in \mathbb{R}$ arbitrarily small and/or $\beta \in \mathbb{R}$ arbitrarily large. 
\end{defn}

We now define the problem of interest in our paper. 

\begin{problem}
\label{prob:1}
In our paper we aim to develop a distributed optimal allocation algorithm for nodes $v_j \in \mathcal{V}_p$ that wish to preserve their privacy when they exchange quantized information with neighboring nodes while calculating $w_j^*$ which fulfills \eqref{cond:balance}. 
Furthermore, nodes need to (i) converge to the optimal solution after a finite number of time steps, (ii) process and transmit quantized values, and (iii) cease transmissions once convergence has been achieved. 
\end{problem}

\section{Distributed Test Kit Allocation With Privacy-Preservation}\label{sec:alloc_alg_priv_sec}

\subsection{Quantized Privacy-Preserving Strategy}\label{quant_steategy}

During the operation of our algorithm we aim to calculate $w_j^*$ which fulfills \eqref{cond:balance} while preserving the privacy of the nodes following the privacy preserving strategy. 
In the current literature (e.g., \citep{4448699, manitara2013QAC, Mo-Murray:2017, 2019:Allerton_themis} and references therein), each node initially injects a nonzero offset to its initial state. 
During the operation of our algorithm, if a node follows the proposed privacy preserving strategy it assigns an offset to each outgoing link (note that the sum of offsets is equal to its initial state). 
Then, if it performs a transmission towards an out-neighbor, it injects the assigned offset to the transmitted variables. 
More specifically, each node $v_j \in \mathcal{V}_p$ maintains a set of values $\text{off}^{(z)}_{lj} \neq 0$, $\text{off}^{(y)}_{lj} \neq 0$, for every $v_l \in \mathcal{N}_j^+$. 
The sum of these values is equal to the node's initial state (i.e., $\sum_{v_l \in \mathcal{N}_j^+} \text{off}^{(z)}_{lj} = l_j + u_j$, and $\sum_{v_l \in \mathcal{N}_j^+} \text{off}^{(y)}_{lj} = \lambda_j$). 
Furthermore, each node $v_j \in \mathcal{V}_p$ maintains a set of counters $\text{off}_{lj} = 1$, for every $v_l \in \mathcal{N}_j^+$ in order to remember whether it has injected every offset to the transmitted messages. 
Then, if node $v_j$ performs a transmission towards out-neighbor $v_l \in \mathcal{N}_j^+$, it injects $\text{off}^{(z)}_{lj}$ and $\text{off}^{(y)}_{lj}$ to the transmitted messages $z_j$, $y_j$, respectively. 
Finally, note that the nodes $v_l \notin \mathcal{V}_p$, either execute the proposed algorithm or execute the quantized average consensus algorithm in \citep{2021:Rikos_Hadj_Johan}. 

\subsection{Optimal Allocation Algorithm with Privacy-Preserving Guarantees}\label{quant_algorithm}


We now present the distributed algorithm (detailed below as Algorithm~\ref{algorithm1}) which solves Problem~\ref{prob:1} described in Section~\ref{Opt_Prob_Form}. 
In order to solve Problem~\ref{prob:1} we need to consider the following assumptions. 

\begin{assum}
\label{assm:connected}
The communication network is modelled as a strongly connected digraph. 
\end{assum}
\begin{assum}
\label{assm:diameter}
An upper bound $D'$ of the diameter $D$ (i.e., $D' \geq D$) is known to every node in the network. 
\end{assum}

Assumption~\ref{assm:connected} ensures that information transmitted by one node can reach every other node, and is important for guaranteeing convergence to the optimal solution. 
Assumption~\ref{assm:diameter}, is required for terminating the operation of Algorithm~\ref{algorithm1} once convergence has been achieved. 

\begin{varalgorithm}{1}
\caption{Quantized Test Kit Allocation Algorithm With Privacy-Preservation}
\noindent \textbf{Input:} A strongly connected digraph $\mathcal{G}_d = (\mathcal{V}, \mathcal{E})$ with $n=|\mathcal{V}|$ nodes and $m=|\mathcal{E}|$ edges. 
Each node $v_j\in \mathcal{V}$ has knowledge of $l_j, u_j, D, \lambda_j \in \mathbb{Z}$. \\
\textbf{Initialization:} Each node $v_j \in \mathcal{V}$ does the following: 
\begin{list4}
\item[$1)$] Assigns a nonzero probability $b_{lj}$ to each of its outgoing edges $m_{lj}$, where $v_l \in \mathcal{N}^+_j \cup \{v_j\}$, as follows
\begin{align*}
b_{lj} = \left\{ \begin{array}{ll}
         \frac{1}{1 + \mathcal{D}_j^+}, & \mbox{if $l = j$ or $v_{l} \in \mathcal{N}_j^+$,}\\
         0, & \mbox{if $l \neq j$ and $v_{l} \notin \mathcal{N}_j^+$.}\end{array} \right. 
\end{align*} 
\item[$2)$] Sets $z_j[0] := \lambda_j$, $y_j[0] = l_j + u_j$, and $\text{flag}_j = 0$.
\item[$3)$] Sets $\text{off}^{(z)}_{lj} \neq 0$, for every $v_l \in \mathcal{N}_j^+$, such that $\sum_{v_l \in \mathcal{N}_j^+} \text{off}^{(z)}_{lj} = l_j + u_j$, and $\text{off}^{(y)}_{lj} \neq 0$, for every $v_l \in \mathcal{N}_j^+$, such that $\sum_{v_l \in \mathcal{N}_j^+} \text{off}^{(y)}_{lj} = \lambda_j$. 
\item[$4)$] Sets $\text{off}_{lj} = 1$, for every $v_l \in \mathcal{N}_j^+$. 
\end{list4} 
\textbf{Iteration:} For $k=1,2,\dots$, each node $v_j \in \mathcal{V}$, does the following: 
\begin{list4} 
\item \textbf{while} $\text{flag}_j = 0$ \textbf{then} 
\begin{list4a}
\item[$1)$] \textbf{if} $k \mod D = 1$  \textbf{then} sets $M_j = \lceil y_j[k] / z_j[k] \rceil$, $m_j = \lfloor y_j[k] / z_j[k] \rfloor$; 
\item[$2)$] \textbf{if} $\sum_{v_l \in \mathcal{N}_j^+} \text{off}_{lj} > 0$  \textbf{then} sets $M_j = M_j + 2$; 
\item[$3)$] broadcasts $M_j$, $m_j$ to every $v_{l} \in \mathcal{N}_j^+$; 
\item[$4)$] receives $M_i$, $m_i$ from every $v_{i} \in \mathcal{N}_j^-$; 
\item[$5)$] sets $M_j = \max_{v_{i} \in \mathcal{N}_j^-\cup \{ v_j \}} \ M_i$, $m_j = \min_{v_{i} \in \mathcal{N}_j^-\cup \{ v_j \}} \ m_i$; 
\item[$6)$] \textbf{if} $z_j[k] > 1$, \textbf{then} calls Algorithm~\ref{algorithm1a}; 
\item \textbf{else if} $z_j[k] \leq 1$, sets $c^{y}_{jj}[k] = y[k]$, $c^{z}_{jj}[k] = z[k]$; 
\item[$7)$] receives $c^{y}_{ji}[k]$, $c^{z}_{ji}[k]$ from $v_i \in \mathcal{N}_j^-$ and sets 
\begin{equation}\label{no_del_eq_y_no_oscil}
y_j[k+1] = c^{y}_{jj}[k] + \sum_{v_i \in \mathcal{N}_j^-}  w_{ji}[k] \ c^{y}_{ji}[k] ,
\end{equation} 
\begin{equation}\label{no_del_eq_z_no_oscil}
z_j[k+1] = c^{z}_{jj}[k] + \sum_{v_i \in \mathcal{N}_j^-} w_{ji}[k] \ c^{z}_{ji}[k] ,
\end{equation}
where $w_{ji}[k] = 1$ if node $v_j$ receives $c^{y}_{ji}[k]$, $c^{z}_{ji}[k]$ from $v_i \in \mathcal{N}_j^-$ at iteration $k$ (otherwise $w_{ji}[k] = 0$); 
\item[$8)$] \textbf{if} $k \mod D = 0$ \textbf{then}, \textbf{if} $M_j - m_j \leq 1$ \textbf{then} sets 
$w_j^* = \lceil \lambda_j  q^s_j[k] \rceil$ and $\text{flag}_j = 1$. 
\end{list4a}
\end{list4}
\textbf{Output:} \eqref{cond:balance} is fulfilled for every $v_j \in \mathcal{V}$. 
\label{algorithm1}
\end{varalgorithm}

\begin{varalgorithm}{1A}
\caption{Quantized Averaging and Offset Injection}
\noindent \textbf{Input:} $z_j[k], y_j[k]$, $z^s_j[k], y^s_j[k]$, $\text{off}^{(y)}_{lj}, \text{off}^{(z)}_{lj}$, $\text{off}_{lj}$, for every $v_l \in \mathcal{N}^+_j$. \\
\textbf{Iteration:} Each node $v_j \in \mathcal{V}$, does the following: 
\begin{list4a}
\item[$1)$] sets $z^s_j[k] = z_j[k]$, $y^s_j[k] = y_j[k]$, 
$
q^s_j[k] = \Bigl \lceil \frac{y^s_j[k]}{z^s_j[k]} \Bigr \rceil \ ;
$
\item[$2)$] sets (i) $mas^y[k] = y_j[k]$, $mas^z[k] = z_j[k]$; (ii) $c^{y}_{lj}[k] = 0$, $c^{z}_{lj}[k] = 0$, for every $v_l \in \mathcal{N}^+_j \cup \{v_j\}$; (iii) $\delta = \lfloor mas^y[k] / mas^z[k] \rfloor$, $mas^{rem}[k]= y_j[k] - \delta \ mas^z[k]$; 
\item[$3)$] \textbf{while} $mas^z[k] > 1$, \textbf{then} 
\begin{list4a} 
\item[$3a)$] chooses $v_l \in \mathcal{N}^+_j \cup \{v_j\}$ randomly according to $b_{lj}$; 
\item[$3b)$] sets (i) $c^{z}_{lj}[k] := c^{z}_{lj}[k] + 1$, $c^{y}_{lj}[k] := c^{y}_{lj}[k] + \delta$; (ii) $mas^z[k] := mas^z[k] - 1$, $mas^y[k] := mas^y[k] - \delta$. 
\item[$3c)$] If $mas^{rem}[k] > 1$, sets $c^{y}_{lj}[k] := c^{y}_{lj}[k] + 1$, $mas^{rem}[k] := mas^{rem}[k]- 1$; 
\end{list4a} 
\item[$4)$] sets $c^{y}_{jj}[k] := c^{y}_{jj}[k] + mas^y[k]$, $c^{z}_{jj}[k] := c^{z}_{jj}[k] + mas^z[k]$; 
\item[$5)$] \textbf{if} $\text{off}_{lj} = 1$, \textbf{then} sets $c^{y}_{lj}[k] = c^{y}_{lj}[k] + \text{off}^{(y)}_{lj}$, $c^{z}_{lj}[k] = c^{z}_{lj}[k] + \text{off}^{(z)}_{lj}$, and $\text{off}_{lj} = 0$; 
\item[$6)$] for every $v_l \in \mathcal{N}^+_j$, if $c^{z}_{lj}[k] > 0$ transmits $c^{y}_{lj}[k]$, $c^{z}_{lj}[k]$ to out-neighbor $v_l$; 
\end{list4a} 
\textbf{Output:} $z^s_j[k], y^s_j[k], q^s_j[k]$, $\text{off}^{(y)}_{lj}$, $\text{off}^{(z)}_{lj}$, $\text{off}_{lj}$. 
\label{algorithm1a}
\end{varalgorithm}

The intuition behind Algorithm \ref{algorithm1} is the following. 
Initially, each node in the set $\mathcal{V}_p$ calculates a set of offsets; one offset for each out-neighbor. 
Then, each node executes the quantized average consensus algorithm in \citep{2021:Rikos_Hadj_Johan}. 
During the operation of the quantized average consensus algorithm, if one node in the set $\mathcal{V}_p$ performs a transmission towards an out-neighbor, it injects the calculated offset to the transmitted variables. 
Finally, if one node in the set $\mathcal{V}_p$ has not transmitted every offset to each out-neighbor, it delays the distributed stopping protocol until every offset is transmitted.  

Note here that every node in the set $\mathcal{V}_p$ that wants to preserve its privacy executes Algorithm~\ref{algorithm1}. 
The set of neutral nodes in $\mathcal{V}_n$ executes the quantized average consensus algorithm in \citep{2021:Rikos_Hadj_Johan}. 
Finally, the set of curious nodes in $\mathcal{V}_c$, either executes Algorithm~\ref{algorithm1} or the quantized average consensus algorithm in \citep{2021:Rikos_Hadj_Johan} (this means that $\mathcal{V}_p$ and $\mathcal{V}_c$ are not necessarily disjoint). 


Next, we show that
Algorithm~\ref{algorithm1} solves Problem~\ref{prob:1} in Section~\ref{Opt_Prob_Form}. 
Due to space limitations we provide a sketch of the proof. 

\begin{thm}\label{thm:test_allocation_private}
Consider a strongly connected digraph $\mathcal{G}_d = (\mathcal{V}, \mathcal{E})$
under Assumptions~\ref{assm:connected}, \ref{assm:diameter}. 
Every node in the set (i) $\mathcal{V}_p$ executes Algorithm~\ref{algorithm1}, (ii) $\mathcal{V}_n$ executes the algorithm in \citep{2021:Rikos_Hadj_Johan}, and (iii) $\mathcal{V}_c$, either executes Algorithm~\ref{algorithm1} or the algorithm in \citep{2021:Rikos_Hadj_Johan}. 
Algorithm~\ref{algorithm1} solves Problem~\ref{prob:1}.
\end{thm}

\textit{Proof:}
The main idea of this proof is that we will calculate (i) the number of time steps in order for every $v_j \in \mathcal{V}_p$ to complete the privacy preservation mechanism (i.e, to inject all its offsets in the network), and (ii) the number of time steps for the algorithm in \citep{2021:Rikos_Hadj_Johan} to converge. 

The operation of Algorithm~\ref{algorithm1} can be interpreted as the ``random walk'' of $2 \lambda_{\mathrm{tot}} - n$ ``tokens'' in a Markov chain (where $\lambda_{\mathrm{tot}} = \sum_{v_{j} \in \mathcal{V}} \lambda_j$ and $n=|\mathcal{V}|$). 
Furthermore, every node has one stored token which is stationary (i.e., it does not perform a random walk). 
Each token contains a pair of values $y[k] \in \mathbb{N}$, $z[k] = 1$. 
Each time two or more tokens meet at a specific node, their $y[k]$ values either become equal or have difference equal to one. 

From  \citep[Lemma~1]{rikos2021optimal} we have that the probability $P^{D+1}_{T^{out}}$ that ``the specific token $T_{\lambda}^{out, \vartheta}$ is at node $v_i$ after $D+1$ time steps, and node $v_i$ transmits to a specific $v_{i'} \in \mathcal{N}^+_i$'' is 
\begin{equation}\label{lowerProf_no_oscil}
P^{D+1}_{T^{out}} \geq (1+\mathcal{D}^+_{max})^{-(D+1)} .
\end{equation}
This means that the probability $P^{D+1}_{N\_ T^{out}}$ that ``the specific token $T_{\lambda}^{out, \vartheta}$ has not visited node $v_i$ after $D+1$ time steps (or has visited but not been transmitted to the specific node $v_{i'} \in \mathcal{N}^+_i$)'' is
\begin{equation}\label{lowerProf_not_no_oscil}
P^{D+1}_{N\_ T^{out}} \leq 1- (1+\mathcal{D}^+_{max})^{-(D+1)} .
\end{equation}

By extending this analysis, we can state that for any $\epsilon$, where $0 < \varepsilon < 1$ and after $\tau(D+1)$ time steps where
\begin{equation}\label{windows_for_conv_1_no_oscil}
\tau \geq \Big \lceil \dfrac{\log{\epsilon}}{\log{(1 - (1+\mathcal{D}^+_{max})^{-(D+1)})}} \Big \rceil ,
\end{equation}
the probability $P^{\tau}_{N\_ T^{out}}$ that ``the specific token $T_{\lambda}^{out, \vartheta}$ has not visited node $v_i$ after $\tau (D+1)$ time steps (or has visited but not been transmitted to the specific node $v_{i'} \in \mathcal{N}^+_i$)'' is
\begin{equation}\label{ProbNotMeet_after_t_no_oscil}
P^{\tau}_{N\_ T^{out}} \leq [P^{D+1}_{N\_ T^{out}}]^{\tau} \leq \epsilon .
\end{equation}
This means that after $\tau (D+1)$ time steps, where $\tau$ fulfills \eqref{windows_for_conv_1_no_oscil}, the probability that ``the specific token $T_{\lambda}^{out, \vartheta}$ has visited node $v_i$ after $\tau (D+1)$ time steps and has been transmitted to a specific $v_{i'} \in \mathcal{N}^+_i$'' is equal to $1-\epsilon$. 
Thus, by extending this analysis, for $k \geq (\mathcal{D}^+_{max}) \tau (D+1)$ we have that every node $v_i$ will perform a transmission towards \textit{every} out-neighbor $v_{i'} \in \mathcal{N}^+_i$ with probability $(1-\epsilon)^{(\mathcal{D}^+_{max})}$. 

Once every node $v_i$ performs a transmission towards \textit{every} out-neighbor $v_{i'} \in \mathcal{N}^+_i$, the privacy preserving strategy has been completed, and the operation of Algorithm~\ref{algorithm1} is similar to \citep{rikos2021optimal}. 
As a result, for the operation of Algorithm~\ref{algorithm1} during time steps $k \geq (\mathcal{D}^+_{max}) \tau (D+1)$ the rest of the proof is similar to Theorem~$3$ in \citep{rikos2021optimal} (since the operation of Algorithm~\ref{algorithm1} for time steps $k \geq (\mathcal{D}^+_{max}) \tau (D+1)$ is identical to \citep{rikos2021optimal}). 

\subsection{Topological Conditions for Privacy Preservation}

We now present, in the following theorem, the necessary topological conditions for privacy preservation. 

\begin{thm}
    \label{thm:privacy_preservation}
    Consider a fixed strongly connected digraph $\mathcal{G}_d = (\mathcal{V}, \mathcal{E})$ under Assumptions~\ref{assm:connected}, \ref{assm:diameter}. Every node in the set (i) $\mathcal{V}_p$ executes Algorithm~\ref{algorithm1}, (ii) $\mathcal{V}_n$ executes the algorithm in \citep{2021:Rikos_Hadj_Johan}, and (iii) $\mathcal{V}_c$, either executes Algorithm~\ref{algorithm1} or the algorithm in \citep{2021:Rikos_Hadj_Johan}. 
    No subset of curious nodes $\mathcal{V}_c$ is able to identify the initial state $y_j[0]$ of $v_j$, if, and only if, the following conditions are fulfilled:
    \begin{enumerate}[label=\roman*)]
        \item $v_j$ has at least one out-neighbor (or in-neighbor) $v_{l}\in \mathcal{V}_p \setminus (\mathcal{V}_c \cup \{ v_j \})$, 
        \item there is a message exchange between $v_j$ and $v_{l}$ while both are implementing the privacy-preserving mechanism, and
        \item $v_{l}$ transmits to an out-neighbor $v_{\ell}$ for the first time during the next time step. 
    \end{enumerate}
\end{thm}

\textit{Proof:}
The proof 
consists of two parts. 
In the first part, we analyze the sufficiency of the above conditions (i) - (iii), and in the second part we analyze their necessity. 

Regarding the first part, let us assume that conditions (i) - (iii) hold.  
Let us assume that nodes $v_j$ and $v_{l}$ are executing Algorithm~\ref{algorithm1} (i.e., $v_j, v_{\ell} \in \mathcal{V}_p$). 
Now let us assume that at time step $k'$, node $v_j$ transmits a message to its out-neighbor $v_{l}$ (the case $v_{l} \in \mathcal{N}_j^-$ can be proven identically). 
Node $v_j$ will inject $\text{off}^{(y)}_{lj}$, $\text{off}^{(z)}_{lj}$ to the transmitted values. 
The values $\text{off}^{(y)}_{lj}$, $\text{off}^{(z)}_{lj}$ are only known to $v_j$ and perhaps to $v_{l}$. 
Then, in the next time step, node $v_{l}$ will transmit to a an out-neighbor $v_{\ell}$ for the first time. 
This means that $v_{l}$ will inject $\text{off}^{(y)}_{\ell l}$, $\text{off}^{(z)}_{\ell l}$ to the transmitted values. 
As a result, the transmitted message depends on the sum of offsets $\text{off}^{(y)}_{lj} + \text{off}^{(y)}_{\ell l}$, and $\text{off}^{(z)}_{lj} + \text{off}^{(z)}_{\ell l}$. 
Since, both $v_j, v_{l} \in \mathcal{V}_p$, the curious nodes may be able to determine $\text{off}^{(y)}_{lj} + \text{off}^{(y)}_{\ell l}$, and $\text{off}^{(z)}_{lj} + \text{off}^{(z)}_{\ell l}$, but not each $\text{off}^{(y)}_{lj}$, $\text{off}^{(y)}_{\ell l}$, and $\text{off}^{(z)}_{lj}$, $\text{off}^{(z)}_{\ell l}$. 
As a result, the privacy of both node $v_j$ and node $v_{l}$ is preserved. 

Regarding the second part, let us assume that condition (i) does not hold. 
In this case, all the in- and out-neighbors of node $v_j$ are curious and they collude with each other. 
This means that the curious nodes will know all the values node $v_j$ transmitted to its out-neighbors, and they will know all the values $v_j$ received from its in-neighbors. 
Therefore, it is not possible for node $v_j$ to keep its privacy. 
Let us now assume that condition (ii) does not hold. 
In this case, non of the in- or out-neighbors of node $v_j$ will inject any offsets to the messages they transmit. 
This means that the curious nodes will know that the transmitted values have only the injected offsets from node $v_j$. 
Therefore, it is not possible for node $v_j$ to keep its privacy. 
Finally, the case where condition (iii) does not hold, the claim
can be proven analogously. 

\section{Simulation Results}\label{simul_results}


\begin{figure}[!htp]
    \centering
    \subfloat[10  nodes\label{fig:2a}]{
      \includegraphics[width=1\columnwidth]{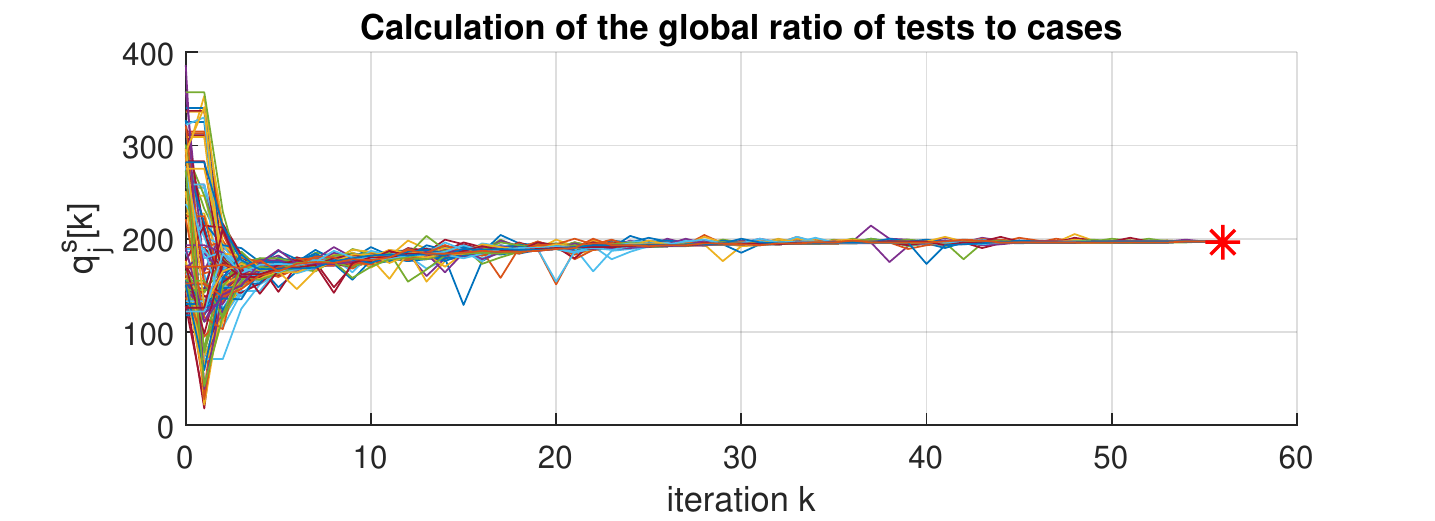}}
    \hfill
    \subfloat[100 nodes\label{fig:2b}]{
      \includegraphics[width=1\columnwidth]{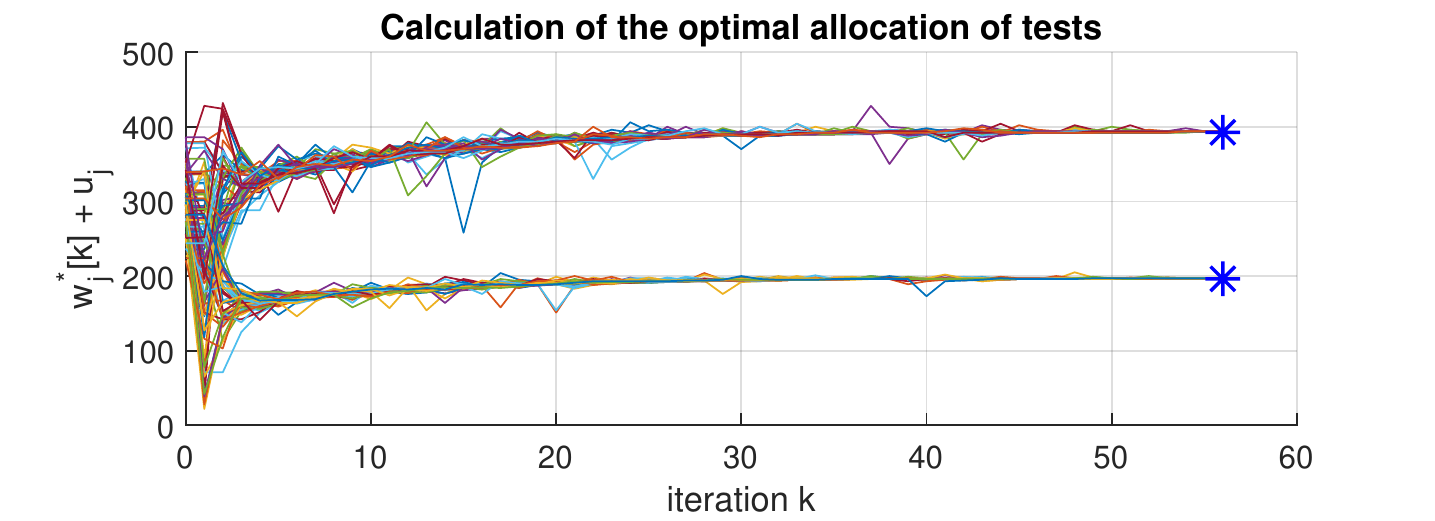}}
    \caption{Convergence towards the global ratio of test kits to infections marked as a red star (\textcolor{red}{*}) (in (a)) and optimal allocation marked as blue stars (\textcolor{blue}{*}) (in (b)) in a random strongly connected network of $100$ nodes during the operation of Algorithm~\ref{algorithm1}. 
    Each line represents the state variable $q_j^s[k]$ in the top plot and $w_j^*[k] + u_j = q_j^s[k] \lambda_j$ in the bottom plot for each $v_j \in \mathcal{V}$. }
  \label{fig:small_lambda}
\end{figure}


\begin{figure}[!htp]
    \centering
    \subfloat[10  nodes\label{fig:3a}]{
      \includegraphics[width=1\columnwidth]{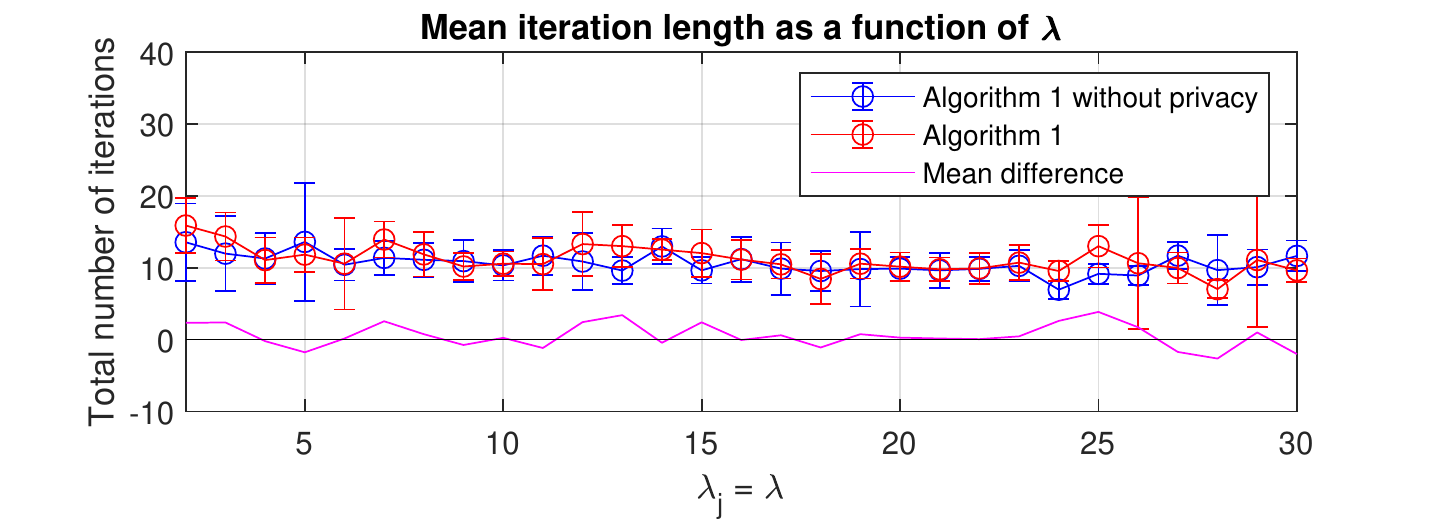}}
    \hfill
    \subfloat[100 nodes\label{fig:3b}]{
      \includegraphics[width=1\columnwidth]{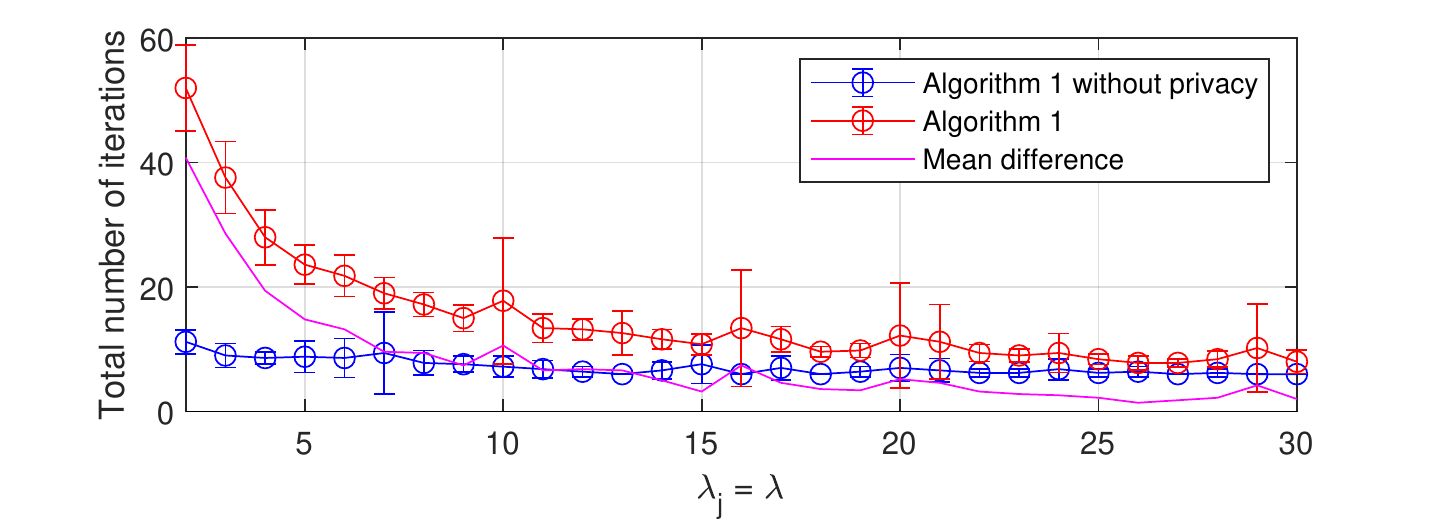}}
    \caption{Comparison between the number of iterations required for Algorithm~\ref{algorithm1}  and the algorithm in \citep{2021:Rikos_Hadj_Johan} to converge over a random network of $10$ nodes (in (a)) and $100$ nodes (in (b)) when $\lambda_j = \lambda$ is set equal for all $v_j \in \mathcal{V}$ and varied. 
    The red and the blue line represents each algorithm, while the pink line is the difference in iteration length. 
    Each data point is the mean iteration length of $100$ simulations (in (a)) and $10$ simulations (in (b)). The error bars correspond to the standard deviation.}
\label{fig:comp_nodes}
\end{figure}

We now illustrate the efficiency of Algorithm~\ref{algorithm1}. 
To this end, we consider the setting where test kits need to be optimally allocated proportionally to the number of infections. 
The operation of Algorithm~\ref{algorithm1} is demonstrated in Fig.~\ref{fig:small_lambda} and Fig.~\ref{fig:comp_nodes}. 
In these figures we show its rate of convergence and the mean number of iterations required for convergence with and without the privacy preservation mechanism, respectively. 

In Figure~\ref{fig:small_lambda} we demonstrate the convergence of Algorithm \ref{algorithm1}. 
Random choices of the initial number of test kits and infections were made such that $l_j + u_j \in [200,400]$, and $\lambda_j = 1 \text{ or } 2$ for all $v_j \in \mathcal{V}$ (which explains the convergence in Fig.~\ref{fig:2b}). 
In Fig.~\ref{fig:2a} each line represents the state variable $q_j^s[k]$ for every iteration step. 
The value of $q_j^s[k]$ corresponds to the calculated value of the global ratio $q$ in \eqref{eq:global_average} at each node for every iteration step. 
In Fig.~\ref{fig:2a} we have that all state variables have converged to either the ceiling or the floor of $q$.
In Fig.~\ref{fig:2b} we show the optimal allocation of test kits proportionally to the number of infections.
Each line is represented by $w_j^* + u_j = q_j^s[k]\lambda_j$ at every iteration step. 
In Fig.~\ref{fig:2b}, the privacy preservation mechanism can be seen as ``spikes'' extending from the lines of Figure \ref{fig:small_lambda}. 
These ``spikes'' denote the offset injection during Iteration Step~$5$ of Algorithm~\ref{algorithm1a}. 

In Figure~\ref{fig:comp_nodes} we show the mean number of iterations required for convergence of Algorithm~\ref{algorithm1} with and without privacy preservation guarantees (i.e., if we execute Algorithm~\ref{algorithm1} or the quantized average consensus algorithm in \citep{2021:Rikos_Hadj_Johan}). 
We consider networks of $10$ and $100$ nodes, and assume that these are strongly connected networks. 
We implement Algorithm~\ref{algorithm1} for both these networks; see Fig.~\ref{fig:3a} and Fig.~\ref{fig:3b}, respectively. 
The number of test kits $l_j + u_j$ is randomly set in the interval $[500, 1500]$ at each node. 
In Fig.~\ref{fig:3a}, both algorithms (i.e., with and without the privacy preservation mechanism) require the same number of time steps for convergence. 
The same holds for Fig.~\ref{fig:3b} for $\lambda_j$ greater than $15$ infections. 
However, in Fig.~\ref{fig:3b}, we have that Algorithm~\ref{algorithm1} requires more time steps for convergence when $\lambda_j \in \{ 1, 2, ..., 10 \}$. 
Note here that in practical scenarios, we would most likely wish to find the optimal allocation of test kits when cities in a country experience more than $15$ cases. 
This means that as long as the considered network is not much greater than $100$ cities, every city may preserve its privacy without any noticeable loss in computation time.

\section{Conclusions and Future Directions}\label{future_dir}

In this paper, we presented a novel distributed privacy-preserving algorithm that optimally allocates test kits proportionally to the number of infections. 
We showed that all nodes calculate the optimal allocation of test kits with high probability after a finite number of time steps while exchanging quantized values. 
Furthermore, once convergence has been achieved every node terminates its operation. 
We also provided sufficient topological conditions for privacy-preservation. 
Finally we presented simulation results of our proposed distributed algorithm, and we demonstrated its convergence rate for networks of various sizes. 

In the future, we plan to extend our algorithm to also handle errors in the interaction messages transmitted between nodes. 

\bibliography{References_KTH_Intern_2021}

\end{document}